\documentclass[10pt,twocolumn]{article}
\usepackage[cp1251]{inputenc}
\usepackage[english]{babel}
\usepackage{graphicx}
\title{{\bf \Large Cosmology with an Effective $\Lambda$-Term in Lyra  Manifold}\\
{\normalsize ~~{\bf V.\,K. Shchigolev}\thanks{E-mail:
vkshch@yahoo.com}}\\
{\small {\it Department of Theoretical Physics, Ulyanovsk State University, Ulyanovsk 432000, Russia}}\\
\vspace{2mm}
\small \begin{quote}{{\it A cosmological model in  Lyra's geometry are studied under the assumption that an effective cosmological term is appeared in the field equations as the result of interaction between the  displacement vector field and an auxiliary $ \Lambda $ -  term.  Some exact solutions of the model equations are obtained and preliminary studied for the simplest cases in order to illustrate how such a model works.}
 \\
\vspace{2,5mm}
{\it PACS: 98.80.-k; 98.80.Jk; 04.20.Jb}\vspace{-1.2cm}}
\end{quote}}
\date{}

\begin{document}

\maketitle

The Lyra's geometry \cite{Lyra} provides  one of the possible alternatives in  modification of the cosmological models, the need for which is almost generally recognized. Such a modification of the gravitational theory has long been known, but now it again attracts attention  due to the opening of the late-time cosmological acceleration. This fact followers from  the supernovae of type Ia observations \cite{Riess, Perlmutter}, Cosmic Microwave Background Radiation \cite{Spergel, Komatsu}, and Baryon Acoustic Oscillations in galaxy surveys \cite{Blake, Seo}. Different versions of the scalar-tensor theories are involved to solve the problem of late acceleration.  Besides, a substantial number of investigations of the models with a varying in time cosmological term has been proposed during the last two decades (see, e.g.  \cite{Chen} -\cite{Carneiro}).

Recently, a number of interesting results have been obtained for the effective cosmological term in Lyra geometry \cite{Hova}-\cite{Shchigolev2}.
One should recall that the displacement vector field has a purely geometrical origin. Therefore, it seems rather logical that such a field in the dynamical equations of cosmological models should be considered as a constant, possibly associated with a cosmological constant, or as a hidden parameter which is masked in some  dynamical parameters of the model. It is this approach to the field of displacement vector is proposed in our work. Generally speaking, such an idea is not something absolutely new. It is involved  in many studies of the models with variable $G$ and decaying vacuum energy density. We intend to apply this idea to the models in Lyra geometry. Initially, an auxiliary cosmological term as a kind of  "bare" cosmological term $\Lambda(t)$, somehow interacting with the displacement vector field, is substituted  into  the gravitational equations.  The result of this interaction is to preserve the continuity equation for the ordinary matter in its standard form, and the generation of a new effective cosmological term. The appearance of the latter occurs  entirely due  to the displacement field. Exactly this  $\Lambda_{eff}(t)$ - term is generated by the displacement field. It determines, along with matter, the dynamics of the cosmological evolution and becomes a constant in its absence, as it should be in the case of GR. We used such a method of generating an effective cosmological term in the context of other cosmological theory in \cite{Shchigolev3}.

The Einstein's field equations in Lyra's geometry, proposed in \cite{Sen} in normal gauge, can be written with a time-varying $\Lambda$ - term as
\begin{equation}\label{1}
R_{ik}- \frac{1}{2} g_{ik} R - \Lambda g_{ik} +  \frac{3}{2}\phi_i \phi_k - \frac{3}{4}g_{ik}\phi^j \phi_j = T_{ik},
\end{equation}
where $\phi_i$ is a displacement vector. For simplicity, we assume  that the gravitational constant $8\pi G=1$. All other symbols have their usual meanings in the Riemannian geometry.
The energy-momentum tensor (EMT) of matter $T_{ik}$ can be derived in a usual manner from the Lagrangian of matter. Considering the matter to be a perfect fluid, we have
\begin{equation}\label{2}
T_{ik}= (\rho_m +p_m)u_i u_k -p_m\, g_{ik},
\end{equation}
where  $u_i = (1,0,0,0)$ is  4-velocity of the co-moving observer, satisfying $u_i u^i = 1$.
Let $\phi_i$ be a time-like vector field of displacement,
\begin{equation}\label{3}
\phi_i = \left(\frac{2}{\sqrt{3}}\,\beta,0,0,0\right),
\end{equation}
where $\beta = \beta(t)$ is a function of time alone, and the factor $2/\sqrt{3}$ is substituted in order to simplify the writing of all the following equations. The line element of a Friedmann-Robertson-Walker (FRW) is represented by
$$ds^2 = d t^2- a^2 (t)(d r^2+\xi^2 (r)d \Omega^2),$$
where $a(t)$ is a scale factor of the Universe, $\xi(r)=\sin r,r,\sinh r$ in accordance with the  sign of the spatial  curvature $k=+1,0,-1$.
Given this metric and Eqs. (\ref{2}), (\ref{3}), we can reduce the field Eq. (\ref{1}) to the following set of equations:
\begin{eqnarray}
3H^2 + \frac{3 k}{a^2} - \beta^2 = \rho_m +\Lambda,\label{4}\\
2 \dot H + 3H^2 + \frac{k}{a^2} + \beta^2 = -  p_m+\Lambda,\label{5}
\end{eqnarray}
where $H = \dot a/a $ is the Hubble parameter, and an overdot stands for  differentiation with respect to cosmic time $t$.

The continuity equation follows from Eqs. (\ref{4}) and (\ref{5}) as:
\begin{equation}\label{6}
\dot \rho_m + \dot \Lambda + 2 \beta \dot \beta + 3 H \Big[\rho_m + p_m + 2\beta^2 \Big]=0.
\end{equation}

From now on, we consider a spatially flat FRW cosmology with $k=0$.
In this case, one can rearrange the basic equations of the model, Eqs. (\ref{4}) and (\ref{5}), as follows:
\begin{equation}\label{7}
3H^2  = \rho_m +\Lambda + \beta^2,~~
2 \dot H  = - (\rho_m + p_m + 2  \beta^2),
\end{equation}
As can be easily verified, the continuity equation (\ref{6}) follows from the set of equations (\ref{7}) in the same form.

Formally, the displacement vector field can be treated through the EMT in the same way as in the most studies on the Lyra geometry in cosmology. In such a case, the law of conservation for the energy-momentum is accounted for the sum of EMTs \cite{Hova}.  Let us emphasize that the main idea of this work is that the displacement vector field has a geometrical nature, but it can give rise to an effective cosmological term $\Lambda_{eff}$. It means we believe that $\Lambda(t)$ is not an independent dynamical parameter of the model, and it should be excluded from the system of Eqs. (\ref{7}), (\ref{9}) and (\ref{10}). This is what leads to an effective cosmological term via the kinematic (geometric) parameter of the model, namely $H(t)$, but only in the presence of the displacement vector field. As well known, the covariant equation for the EMT $T^k_{i\,; k} = 0$ follows from the identity $G^k_{i\,; k} = 0$ for the Einstein tensor. So, if we are going to preserve the continuity equation for matter in its standard form, the field equation (\ref{1}) yields  the following equation for the $\Lambda$ term and displacement field:
\begin{equation}\label{8}
 \Lambda_{;i} =  \frac{3}{2}\Big(\phi_i \phi^k - \frac{1}{2}\delta_i^k\phi^j \phi_j\Big)_{;k}.
\end{equation}
Consequently, we have the following continuity equation for the EMT of matter (\ref{2}):
\begin{equation}
\dot \rho_m + 3 H (\rho_m + p_m )=0.\label {9}
\end{equation}
Substituting Eq. (\ref{3}) into (\ref{8}) or taking into account  Eq. (\ref{9}) in (\ref{6}) , one can obtain that
\begin{equation}\label{10}
\dot \Lambda + 2 \beta \dot \beta + 6 H \beta^2 =0.
\end{equation}
As this $\Lambda$ - term is absent or equals to a constant, we have $\Lambda_{;\, i}\equiv 0$, and  Eq. (\ref{10}) supposes the displacement field to be the so called stiff fluid. The assumption of a non-vanishing and time-varying $\Lambda$ term gives us some new possibilities \cite{Shchigolev2}.
Here, we study our model with  the assumption of $\Lambda \neq const$. Moreover, we suppose that  Eq. (\ref{10}) is satisfied by a curtain $\Lambda(t)$, which can be found as the result of formal integration of Eq. (\ ref {10}):
$$
\Lambda = -\beta^2-6\int H \beta^2 d t +\Lambda_0,
$$
where $\Lambda_0$ is a constant of integration. The substitution of this expression into Eq. (\ref{7}) allows us to get the basic equations of the model as
\begin{eqnarray}
3H^2  = \rho_{eff},\label{11}\\
2 \dot H  = - (\rho_{eff} + p_{eff}),\label{12}
\end{eqnarray}
where we have introduced the effective cosmological constant
\begin{equation}\label{13}
\Lambda_{eff}(t)=\Lambda_0 - 6 \int H \beta^2 d t,
\end{equation}
and the effective values of the energy density and pressure with the following expressions:
\begin{equation}\label{14}
\rho_{eff}=\rho_m +\Lambda_{eff},\,\,\,\,p_{eff}=p_m -\Lambda_{eff} - \frac{\dot\Lambda_{eff}}{3 H}.
\end{equation}
Therefore, the auxiliary $\Lambda$ - term is disappeared now, and an effective $\Lambda_{eff}$ - term (\ref{13}),  generated by the displacement field, is appeared.

It is easy to verify that due to  Eqs. (\ref{11}), (\ref{12}), and the continuity equation (\ref{9}), the effective energy density and pressure (\ref{14}) also satisfy the continuity equation in its usual form:
\begin{equation}
\dot \rho_{eff} + 3 H (\rho_{eff} + p_{eff} )=0.\label {15}
\end{equation}
Below, we provide three examples of solution for our model in the simple cases.

1.  First of all, we consider the case of inflationary model, when $H=H_0 = const.$ and $a(t)=a_0 \exp(H_0 t)$. Then from Eq. (\ref{13}), we get that
$$\displaystyle \Lambda_{eff}=\Lambda_0 - 6H_0\int \beta^2 d t .$$
According to Eqs. (\ref{11}), (\ref{12}) and (\ref{14}), it is easy to find that
\begin{eqnarray}
  p_m &=&\Lambda_0-3H_0^2-6H_0\int \beta^2 d t -2 \beta^2\,,  \label{16}\\
  \rho_m &=& -\Lambda_0+3H_0^2+6H_0\int \beta^2 d t.\ \label{17}
\end{eqnarray}
In addition, it follows that
\begin{equation}\label{18}
\rho_m + p_m = -2\beta^2,
\end{equation}

As well known, there are several approaches to investigate stability of a model. For instance, one can use the fact that the sound-speed squared should not be negative,
\begin{equation}\label{19}
c^2_s=\frac{\dot p}{\dot \rho} \ge 0.
\end{equation}
Let us apply this relation to our model. Substituting Eqs. (\ref{16}) and (\ref{17}) into Eq. (\ref{18}), we conclude that this stability condition can be expressed as
\begin{equation}\label{20}
c^2_s=-1-\frac{1}{3 H_0}\,\frac{d \ln \beta^2}{dt}\ge 0.
\end{equation}

If we assume that our universe is filled with a perfect fluid described by some effective EoS $p_m = w_m \rho_0$, then we have $w_m = -1 -2\beta^2/\rho_m$. Therefore, due to the real character of the displacement vector and non-negativity of the matter density,  we obtain  $w_m \le -1$. It means that the matter should be of the phantom nature. For example, it can be a phantom scalar field $\phi(t)$, for which $\rho_m=-\dot \phi^2/2 + V(\phi), \, p_m=\rho_m=-\dot \phi^2/2 - V(\phi)$.
From this and Eq. (\ref{18}), it follows that $\dot \phi^2=2\beta^2$. As we have $\rho_m - p_m = 2 V(\phi)$ for the potential of phantom field, it follows from Eqs. (\ref{16}), (\ref{17}) that
\begin{equation}\label{21}
V(\phi) =  \frac{\dot \phi^2}{2} + 3H_0\int  \dot \phi^2 d t+3H_0^2-\Lambda_0.
\end{equation}
In this case, the stability condition (\ref{20}) can be written as: $ \displaystyle \frac{d \ln \dot \phi^2}{d t} \le -3 H_0$. Because the logarithmic function changes  monotonically, we conclude that this model can be represented  only by a phantom field with decreasing kinetic term which satisfies the last inequality. Differentiating Eq. (\ref{21}) with respect to time, we see that the phantom field satisfies the usual equation
\begin{equation}\label{22}
\ddot \phi + 3 H_0 \dot \phi - \frac{d V}{d \phi}=0.
\end{equation}
So for the evolving exponentially  field with $\phi = \phi_0 \exp(- \lambda t)$, we have the potential (\ref{21}) of the form
\begin{equation}\label{23}
V(\phi) = \frac{\lambda}{2}(\lambda-3H_0)\phi^2 + V_0,
\end{equation}
where $V_0$ is a constant, and the stability condition (\ref{20}) reduces to the inequality $\lambda \ge (3/2) H_0$. At this,  the displacement vector field is obtained from Eq. (\ref{18}) in the form: $\beta(t)= \pm (\lambda \phi_0/\sqrt{2})\exp(-\lambda t)$.

In addition, a class of solutions in this case can be obtained in the framework of the so-called method of generating function (see, e.g., \cite{Shchigolev1}). Following this method, we put $\dot \phi = F(\phi)$, where $F(\phi)$ is a continuous function. In this case, the expression (\ref{21}) for the potential can be rewritten as
$$
V(\phi) =  \frac{1}{2}F^2(\phi) + 3H_0\int F(\phi) d \phi+3H_0^2-\Lambda_0,
$$
and the stability condition (\ref{20}) takes the following form:
$$
\frac{d\,F(\phi)}{d\,\phi} \le -\frac{3}{2}H_0.
$$
We can see that the case considered above corresponds to the particular choice of the generating function,  $F(\phi) = - \lambda \phi $. Obviously, this case is far from exhausting all possible solutions of the problem.

2. In the second simplest case when the displacement vector is constant, that is $\beta^2=\beta_0^2$, we have from  (\ref{13}) that
\begin{equation}\label{24}
\Lambda_{eff}(t)=\Lambda_0 - 6 \beta_0^2 \ln [a(t)/a_0],
\end{equation}
where $a_0$ is a constant of integration (the present-day value of the scale factor).

If we assume that the matter component of the Universe can be described by a barotropic perfect fluid with some constant EoS parameter: $w_m \leq 1$, then Eq. (\ref{9}) can be explicitly  integrated as
\begin{equation}\label{25}
\rho_m = \rho_0 a^{\displaystyle -3(1+w_m)},
\end{equation}
where $\rho_0$ is a constant of integration.
With the help of Eqs. (\ref{14}), (\ref{24}) and (\ref{25}), the set of equations (\ref{11}), (\ref{12}) takes the following form:
\begin{equation}\label{26}
3H^2  = \rho_0 a^{\displaystyle -3(1+w_m)} +\Lambda_0 - 6 \beta_0^2 \ln [a(t)/a_0],
\end{equation}
\begin{equation}\label{27}
2 \dot H  = - (1+w_m)\rho_0 a^{\displaystyle -3(1+w_m)} - 2 \beta_0^2.
\end{equation}
It should be noted that Eq. (\ref{27}) is a differential consequence of  Eq. (\ref{26}) that  can be easily verified. Thus, only Eq. (\ ref {26}) is independent and able to determine the dynamics of the system.

As known, the SNIa Union2 database includes 557 SNIa \cite{Amanullah} and provides one of the possible observational restrictions on cosmological models. For this end, we express the scale factor $a$  in terms of redshift $z$ by the relation
\begin{equation}\label{28}
a =\frac{1}{1+z}
\end{equation}
where we put the present value of scale factor to be 1.
Considering the definitions of the dimensionless energy densities,
$\Omega_{m0}=\rho_0/3H_0^2 a_0^{3(1+w_m)},\,\Omega_{\Lambda_0}=\Lambda_0/3H_0^2,\,\Omega_{\beta_0}=2\beta_0^2/H_0^2$
, the Friedmann equation (\ref{26})
can be written as
\begin{eqnarray}\label{29}
E^2(z)=\frac{H^2}{H_0^2}=\Omega_{m0}(1+z)^{3(1+w_m)} \nonumber\\+\Omega_{\Lambda_0} + \Omega_{\beta_0}\ln(1+z),
\end{eqnarray}
where $H_0$ is the present value of the Hubble parameter.
As well known \cite{Chen}, the distance modulus $\mu$ is given by
\begin{equation}\label{30}
\mu(z) = 5 \log_{\,10} (d_L/Mpc) +25
\end{equation}
where $d_L$ is the luminosity distance. In the flat universe, it is connected with redshift as follows
\begin{equation}\label{31}
d_L = (1+z)\int\limits_o^z \frac{d z'}{H_0 E(z')}.
\end{equation}

Let us study Eq. (\ref{29}) for a small value $z$ compared with unity , that is for the epoch close to the present time. Retaining in the series expansion of the power and logarithmic functions only linear terms, we obtain
\begin{equation}\label{32}
E^2(z)\approx (\Omega_{m0}+\Omega_{\Lambda_0}) \nonumber\\
+[3(1+w_m)\Omega_{m0}+\Omega_{\beta_0}]\,z.
\end{equation}
From Eqs. (\ref{31}) and (\ref{32}), and using the relation $\Omega_{m0}+\Omega_{\Lambda_0}=1$ followed from (\ref{29}) as $z=0$, we have
$$
d_L = \frac{2 (1+z)\Big(\sqrt{\vphantom{\dot A} 1+ [3(1+w_m)\Omega_{m0}+\Omega_{\beta_0}]z}-1\Big)}{\vphantom{\dot A}H_0[3(1+w_m)\Omega_{m0}+\Omega_{\beta_0}]}.
$$
Substituting this expression into Eq. (\ref{30}), we obtain  the distance modulus  in terms of redshift parameter $z$.
Following the usual procedure (see, e.g. \cite{Liao} and bibliography therein), we can find the best-fit values of $\Omega_{\beta_0}$ and $\Omega_{m0}$. Thus, one could be able to estimate the magnitude of $\beta_0^2 = H_0^2\Omega_{\beta_0}/2$.

3. As it follows from Eq. (\ref{27}), a wide class of solution for our model can
be obtained on the basis of the phenomenological laws for the evolution of cosmological term, understanding by this its effective value.  At this, we are able to consider many different laws for a time-varying cosmological term, represented in the literature \cite{Sahni}. Indeed, substituting the effective cosmological term   of the form $\Lambda_{eff}(t) = L(t,a,H)$, where $L(t,a,H)$ is a differentiable function, into Eq. (\ref{13}), we obtain after differentiation with respect to time:
\begin{equation}\label{33}
\beta^2 = -\frac{1}{6H}\frac{\partial L}{\partial t} -\frac{1}{6H}\frac{\partial L}{\partial a} \dot a - \frac{\dot H}{6H}\frac{\partial L}{\partial H}.
\end{equation}
This equation can be considered as the main one for searching  $a(t)\Rightarrow \beta(t)$ or $\beta(t)\Rightarrow a(t)$. After that,  the rest parameters of this model can be obtained from Eqs. (\ref{11}), (\ref{12}) and (\ref{14}).
In some cases depending on the specific function $\Lambda_{eff}(t) = L(t,a,H)$ , the solution for $a(t)$ or $H(t)$ can be found from Eq. (\ref{33}), in some cases - even algebraically.  Let us consider just one illustrative example, making use of the phenomenological law $L=\alpha/ a^2(t)$. Then Eq. (\ref{33}) can be reduced to the following one
\begin{equation}\label{34}
\beta^2(t)=\frac{\alpha}{3a^2(t)}.
\end{equation}
Therefore, we obtain from (\ref{14}) and (\ref{25}) that
\begin{equation}\label{35}
\rho_{eff}=\rho_0 a^{\displaystyle -3(1+w_m)} + \Lambda_0 + \alpha a^{-2},
\end{equation}
\begin{equation}\label{36}
p_{eff}= w_m \rho_0 a^{\displaystyle -3(1+w_m)} - \Lambda_0 - \frac{1}{3}\alpha a^{-2},
\end{equation}
From a comparison of Eqs. (\ref{35}),  (\ref{36}) with the set of equations (\ref{11}), (\ref{12}), we can see that the term with $\alpha$ in the effective parameters reveals itself quite similar to the curvature of three-dimensional cross-section $\sim k/a^2$. However, recall that the model is built in the flat space-time with $k = 0$. The appearance of this curvature-like contribution in the energy and pressure is absolutely due to the  displacement vector, as it follows from Eq. (\ref{34}).
\begin{figure}[t]
\centering
\includegraphics[width=0.4\textwidth]{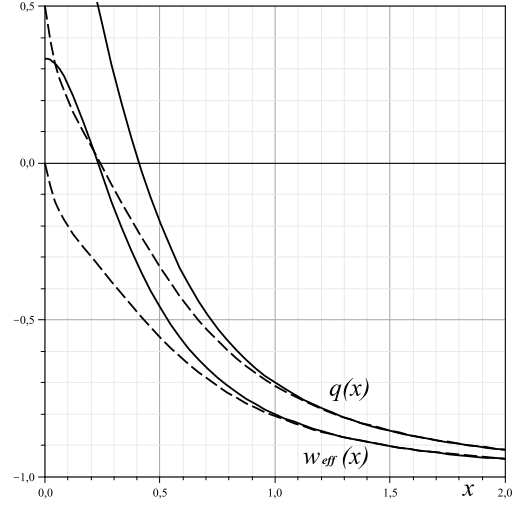}\\
\caption{The plot of effective EoS index $w_{eff}$ and deceleration parameter $q$ versus $x=a/a_0$ for  $w_m=1/3$ (solid line) and $w_m=0$ (dashed line).}
\label{pic1}
\end{figure}
Introducing in addition to $\Omega_{m0}$ and $\Omega_{\Lambda_0}$ one more dimensionless energy density $\Omega_{\alpha} = \alpha/3H_0^2 a_0^{2}$, the following equation for the effective EoS  $w_{eff}=p_{eff}/\rho_{eff}$ can be obtained
\begin{equation}\label{37}
w_{eff}=\frac{w_m \Omega_{m0} - \Omega_{\Lambda_0}x^{\displaystyle 3(1+w_m)}-\displaystyle \frac{1}{3} \Omega_{\alpha}x^{\displaystyle (1+3w_m)}}{\Omega_{m0} + \Omega_{\Lambda_0}x^{\displaystyle 3(1+w_m)}+ \Omega_{\alpha}x^{\displaystyle (1+3w_m)}},
\end{equation}
where and below $x=a/a_0$ is the dimensionless scale factor. At this, the deceleration parameter can be obtained as  $q=-1-\dot H/H^2=(1+3 w_{eff})/2$.
Assuming $w_m > -1$, we have from Eq. (\ref{37})  that  $w_{eff} \to -1$ as $x \to \infty$ regardless of the values $\Omega_{\alpha}$ and $\Omega_{\Lambda_0}$. In Fig.1, the graphs of effective EoS versus the scale factor are presented for two different $w_m$ and the following values of relative densities: $\Omega_{m0}=0.02,\,\Omega_{\Lambda_0}=0.72$ and $\Omega_{\alpha}=0.26$. Note that the effective EoS is able to cross the phantom divide line $w_{eff}=-1$ at a certain value $x = \bar x$, which satisfies  the following equation according to Eq. (\ref{37}):
$$
\bar x^{\displaystyle 1+3 w_m}=-(w_m + 1)\frac{3\, \Omega_{m0}}{2\, \Omega_{\alpha}}.
$$
If $\bar x \ne 0$, then the crossing occurs  at $w_m < -1$.

Embedding  Eq. (\ref{35}) into the Friedmann equation (\ref{11}), we obtain the following equation for the dimensionless scale factor
\begin{equation}\label{38}
H_0^{-2}\, \dot x^2 = \Omega_{m0} x^{\displaystyle -(1+3w_m)}+\Omega_{\Lambda_0} x^2+\Omega_{\alpha}.
\end{equation}
One can find the analytical solutions to this equation for some values of the barometric index of matter $w_m$. Let us note an interesting feature of this equation regarding its coefficients, i.e., the relative densities of energy. Assuming that at the early stages of evolution the matter exists mostly in the form of radiation, i.e. $w_m=w_r =1/3$, then we can rewrite Eq. (\ ref {38}) for a non-negative $y = x^2$ as
\begin{equation}\label{39}
\dot y = 2 H_0 \sqrt{ \Omega_{m0}+\Omega_{\alpha} y+\Omega_{\Lambda_0} y^2}.
\end{equation}
The real and non-oscillating solutions for $y$ can be obtained if the discriminant of quadratic polynomial is not negative: $\Omega_{\alpha}^2-4 \Omega_{m0} \Omega_{\Lambda_0} \ge 0$. Assuming the limiting case of zero value for the latter and taking into account the identity $\Omega_{m0}+\Omega_{\Lambda_0}+\Omega_{\alpha}=1$, we can obtain the following  interesting relations:
\begin{equation}
\!\Omega_{m0}=\Big(1-\sqrt{\Omega_{\Lambda_0}}\,\Big)^2,\,\,
\Omega_{\alpha}=2 \sqrt{\Omega_{\Lambda_0}}\,\Big(1-\sqrt{\Omega_{\Lambda_0}}\,\Big).\label{40}
\end{equation}
Choosing as an example $\Omega_{\Lambda_0}=0.72$ \cite{Liao},  we obtain  from (\ref{40}) the approximate values $\Omega_{m0}=0.02,\,\Omega_{\alpha}=0.26$, which we used above. When these conditions,  Eq. (\ref{40}) can be easily integrated:
\begin{equation}\label{41}
a=a_0\left[\frac{1-\sqrt{\Omega_{\Lambda_0}}}{\sqrt{\Omega_{\Lambda_0}}}\left(e^{\displaystyle 2H_0 \sqrt{\Omega_{\Lambda_0}}t}-1\right)\right]^{1/2},
\end{equation}
where the constant of integration is chosen according to the condition $a=0$ as $t=0$. Because $a_0$ corresponds to the present scale factor, it is easy to find the age of the universe $t_0$ from (\ref{41}) as
$$
t_0 = - \frac{1}{2 H_0 \sqrt{\Omega_{\Lambda_0}}} \ln \left(1-\sqrt{\Omega_{\Lambda_0}}\,\right),
$$
that for $\Omega_{\Lambda_0}=0.72$ gives $t_0 \approx 1.112/H_0 $.

In conclusion, we note that the above study of cosmological models in Lyra geometry with  effective $\Lambda$ term,  generated by the displacement vector, is  preliminary. To make this model more realistic, it is necessary, at least, to introduce one more matter component that interacts with the first one or not. Each of these components should be responsible for its own contribution to the total energy density. Further details and consequences of the models considered
here are in progress.

\end{document}